\documentclass[reprint,amssymb,aps,pre,twocolumn,floats,showpacs]{revtex4-1}
\usepackage{graphicx}
\usepackage{xcolor}
\usepackage{dcolumn}
\usepackage{bm}
\usepackage{amsmath}
\usepackage{soul}

\begin{document}
\title{Characterizing random 1D media with an embedded reflector via scattered waves}
\author{Yiming Huang$^{1}$, Xujun Ma$^{1}$, Azriel Z. Genack$^{1}$, and Victor A. Gopar$^{2,\dagger}$}
\affiliation{$^1$ Department of Physics, Queens College and The Graduate Center of the City University of New York, Flushing, NY, 11367 USA.\\
$^2$ Departamento de F\'isica Te\'orica and BIFI, Universidad de Zaragoza, Pedro Cerbuna 12, E-50009, Zaragoza, Spain.}


\begin{abstract}
We show in random matrix theory, microwave measurements, and computer simulations that the mean free path of a random medium and the strength and position of an embedded reflector can be determined from radiation scattered by the system. The mean free path and strength of the reflector are determined from the statistics of transmission. The statistics of transmission are 
independent of the position of the reflector. The reflector's position can be found, however, from the average 
dwell time for waves incident from one side of the sample. 
\end{abstract}
\maketitle
\section{Introduction}

Since disorder is more the rule than the exception, there is a pressing need in diverse settings to image objects embedded in complex systems. These include medical diagnostics \cite{Jacques,imagingbook}, security inspection, microwave cellular communication, and the search for structural defects in manufactured good, and geological exploration \cite{Etgen,Virieux}. A starting point for the study of the interaction of waves in complex structures is the propagation of waves in uniformly disordered media. This encompasses optical scattering from clouds and paint, electrical resistance in wires, and the spread of Bose-Einstein condensates in random speckle patterns. In random 1D systems, the scaling of the statistics of transport \cite{SPS} and the excitation inside the medium \cite{2017} may be described in terms of a single dimensionless parameter, the ratio of the sample length and the mean free path, $s=L/\ell$ 
\cite{SPS,Mello-book}.

In diffusive or metallic quasi-1D systems with constant cross section, the statistics of propagation may also be given in terms of a single parameter, the ratio of the sample length and the localization length, $L/\xi$. The localization length in quasi-1D media is the product of the number of transverse propagation channels coupled to the sample on either side and the transport mean free path, $\xi=N\ell$ \cite{1977a, 1979a, 1981c}. 

In 1D, it is convenient to study the statistics of the logarithm of transmission since this quantity self-averages \cite{SPS}. For $L \gg \ell$, the statistics of $\ln T$ are log-normal with a variance equal to twice the magnitude of the average value, var$(\ln T)=-2\langle  \ln T \rangle=2L/\ell$. 
In practice, disorder is not uniform throughout a sample. A common example of nonuniformity is the presence of partial reflectors either at the sample’s boundaries  \cite{1991h,Lagendijk,1992,1993b,1994,1997,1999b,2012,Cheng2017,Han2019} or in its interior \cite{Nazarov,Huang}. 

We consider here an ensemble of random configurations with a fixed reflector within the medium. A first step is to describe wave propagation in systems with a reflector in terms of the parameter $s$ of the medium and a set of parameters that characterize the reflector. In 1D, the reflector may be characterized by its flux transmission coefficient, $\Gamma$, and position, $x_0$. An important question is whether it is possible to determine the parameter $s$, as well as $\Gamma$ and $x_0$ from measurements of the statistics of scattered waves. 

Previous work has addressed the challenge of finding the transport mean free path in random multichannel systems in the presence of surface reflection \cite{1991h,Lagendijk,1992,1993b,1994,1997,1999b,2012,Cheng2017}. In samples in which there is a mismatch in the refraction of index across the sample interface, internal scattering is enhanced due to waves making impinging on the interface at angles beyond the critical angle \cite{1991h}.

Scattering of the wave back into the medium at the interface affects the total transmission of an incident wave, \cite{1991h,Lagendijk,1993b}
and the transmittance or optical conductance  \cite{Nazarov,2012}, which is the sum of total transmission over all channels on one side of a sample. The spatial distribution of the wave in the sample \cite{1993b} and the dynamics of scattering \cite{Lagendijk} are also affected by scattering at the interface. It is therefore a challenge to find the values of the mean free path and interfacial reflection. 

The impact of internal reflection can be modeled in terms of boundary extrapolation lengths on the input and output surfaces, and a penetration depth at which the wave direction is assumed to be randomized \cite{1991h,1992,Lagendijk,1993b,1994,1997,1999b,2012}. The situation is simpler in 1D samples without a reflector since the wave can only approach a surface in a single channel so that total internal reflection does not occur. However, reflectors can be present  in 1D media and including a reflector allows for consideration of the broader problem of separately determining the scattering properties of the random medium and of an added fixed element.

In this article, we address the challenge of finding the mean free path of the medium, as well as the 
transmission coefficient $\Gamma$ and position $x_0$ of an embedded reflector in a disordered 1D system from waves scattered from the material. The mean free path and the strength of the reflector can be obtained from the statistics of the transmitted waves. In particular, $\Gamma$ depends on the ensemble averages   
$\langle \ln T \rangle$ and $\langle 1/T \rangle$. The statistical properties of the transmission are independent of the position of the reflector. However, the time delay is sensitive to the reflector's position and can be determined from the average of the time delay of an   incident wave from one side of the sample. Measurements in microwave waveguides and numerical simulations support predictions of random matrix theory (RMT).

\begin{figure}
\begin{center}
\includegraphics[width=1\columnwidth]{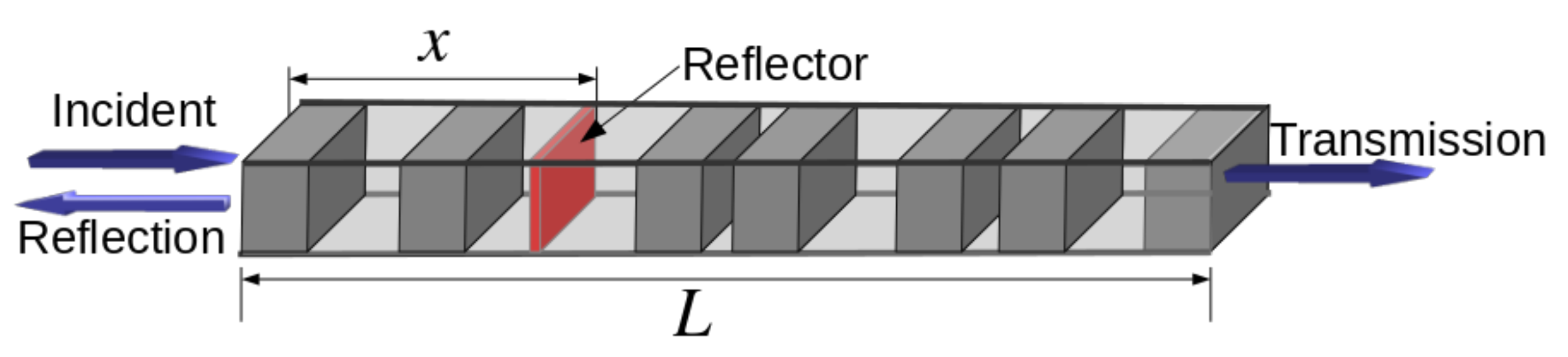}
\caption{Schematic drawing of the random waveguide with a reflector inside.}
\label{fig_1} 
\end{center}
\end{figure}

We carry our microwave experiments in a single-mode waveguide of length $L = 80$ cm composed of randomly selected alternating ceramic slabs and thin-wall Teflon U-channel elements. The ceramic slabs have dielectric constant $\epsilon=3$, height of 22.5 mm, width of 9.8 mm, and thickness of 6.6 mm, and covering $95\%$ percent of the waveguide cross section. The Teflon U-channel spacers have lengths of 1.27, 2.55, 3.82 cm. Successive elements in each sample configuration selected randomly with a probability of 1/6 for the ceramic slabs and 5/18 for each thickness of the U-channel spacers, giving an average of 5 ceramic slabs in each configuration. Measurements are made for four different reflecting copper plates with widths of 22.0 mm and thickness of 0.32 mm  and heights of 9.0, 8.5, 8.0, or 7.0 mm. A diagram of the random sample is shown in Fig \ref{fig_1}.

The wave is launched from one end of the waveguide and the field before and after the sample is detected with an antenna inserted successively into 5 holes before and after the sample. The horizontal motion of the antenna is controlled by a translation stage and the tip is lowered to penetrate to a depth of 0.11 mm below the bottom of the hole. The reflection coefficient $R$ and the incident intensity $I_0$ are found by fitting the intensity of the wave in the segment just before the sample to a constant plus a sinusoidal function. The transmitted wave in the segment right after the sample is normalized by the incident intensity to give the transmission coefficient $T$. 

In order to compensate for absorption within the dielectric, waveguide walls, and for leakage of energy through the holes, we determine the decay rate due to such losses from the linewidth of the narrowest mode in the angular frequency units when a reflecting aluminum plate is placed on the RHS of the sample and the LHS is largely covered by a copper plate to serve as a partial reflector that admits a small fraction of the energy from the source. The decay rate found is $\gamma=9.27\times 10^{-3}  \mathrm{ns}^{-1}$.
The measured spectrum is Fourier transformed into the time domain, multiplied by a factor $e^{\gamma t/2}$ and finally transformed back to the frequency domain. This gives the spectrum that would be obtained without loss. When the correction for absorption is made, the value of $\langle \ln T \rangle$ changes by less than $2\%$ indicating that the sample is much shorter than the absorption length. 

\section{model}
We assume that the complex scattering processes in the random waveguide containing a reflector can be described by the product of three transfer matrices $M_{\rm{l}}$, $M_\Gamma$, and $M_{\rm{r}}$: $M_{\rm{l}}$ is associated to the segment of the waveguide to the left of the reflector, $M_\Gamma$, to the reflector, and $M_{\rm{r}}$, to the segment of the waveguide to the right of the reflector. The transfer matrix $M$ of the entire waveguide is thus given by
\begin{equation}
 M = M_{\rm{r}} M_\Gamma M_{\rm{l}} .
 \label{Mtotal}
\end{equation}

Using the polar representation, the left and right transfer matrices may be written as \cite{Mello-book}
\begin{eqnarray}
\label{MLR}
 M_{\rm{r}(\rm{l})}&& = \nonumber \\
&&
\begin{bmatrix}
   \sqrt{1+\lambda_{\rm{l}(\rm{r})}}e^{i \theta_{\rm{l}(\rm{r})}}   &  \sqrt{\lambda_{\rm{l}(\rm{r})}}e^{i\left (2\mu_{\rm{l}(\rm{r})}-\theta_{\rm{l}(\rm{r})}\right)} \\
    \sqrt{\lambda_{\rm{l}(\rm{r})}}e^{-i\left(2\mu_{\rm{l}(\rm{r})}-\theta_{\rm{l}(\rm{r})}\right)}  & \sqrt{1+\lambda_{\rm{l}(\rm{r})}}e^{-i \theta_{\rm{l}(\rm{r})}}  \nonumber \\
\end{bmatrix}, \\
\end{eqnarray}
where the radial variable $\lambda_{\rm{l}(\rm{r})}$  is a positive real number and $\theta_{\rm{l}(\rm{r})}$  and $\mu_{\rm{l}(\rm{r})}$  are phases.

Assuming that the transmission and reflection coefficients of the reflector are $\Gamma$ and $1-\Gamma$, respectively, with $0\leq \Gamma \leq 1$ and imposing current conservation and time-reversal symmetry, we model the scattering matrix $S_\Gamma$ 
associated with the reflector as
\begin{equation}
S_{\Gamma}=\begin{bmatrix}
    {i\sqrt{1- \Gamma}} &  \sqrt{{\Gamma}} \\
      \sqrt{{\Gamma}}  &  {i\sqrt{1-\Gamma}}
\end{bmatrix}.
\end{equation}
From $S_\Gamma$ matrix, the associated transfer matrix $M_{\Gamma}$  is given by 
\begin{equation}
\label{MGamma}
M_{\Gamma}= \frac{1}{\sqrt{\Gamma}} \begin{bmatrix} 
    1 &  i\sqrt{1-\Gamma} \\
      -i\sqrt{1-\Gamma}  &  1
\end{bmatrix}.
\end{equation}
Since flux is conserved, the transfer matrix $M$ in Eq.~(\ref{Mtotal}) satisfies $\det(M)=1$.

We are interested in the transmission $T$ which is given by the inverse of the element $M_{11}$ of the transfer matrix in Eq.~(\ref{Mtotal}): 
$T=1/M_{11}$. The strength of the reflector will be found from the statistics of $T$. Since transmission statistics are independent of the position of the reflector, \cite{Huang}, we first consider the case with the simplest transfer matrix of a reflector placed at the output of the waveguide. The transfer matrix $M$ in this case is  given by $M=M_{\Gamma}M_{\rm{l}}$. It follows from Eqs. (\ref{MLR}) and (\ref{MGamma}) that 
\begin{equation}
\label{Ttotal}
 \frac{1}{T} =\frac{1}{\Gamma}\left( 1+ (2-\Gamma)\lambda+2 \sqrt{(1-\Gamma)\lambda(1+\lambda)}\cos2\mu \right).
\end{equation}
 The subindex $\rm{l}$ in $\lambda_{\rm{l}}$ and $\mu_{\rm{l}}$ are dropped to simplify the notation.

To calculate the statistical properties of transmission, we assume that $\mu$ in Eq. (\ref{Ttotal}) is uniformly distributed over the interval $(0, 2\pi]$, and $\lambda$ follows the probability density distribution \cite{Molina,Ioannis}
\begin{eqnarray}
\label{poflambda}
 p(\lambda)= \frac{C}{\left(1+\lambda\right)^{1/4}}&&\mathrm{acosh^{1/2}}{\sqrt{1+\lambda}} \nonumber \\
 && \times \exp{\left[-\frac{1}{s} \mathrm{acosh}^2\sqrt{1+\lambda}\right]}, \nonumber \\
\end{eqnarray}
where $C$ is a normalization constant.

\section{Results}
\subsection{Strength of the reflector}

\begin{figure*}[htbp]
\begin{center}
\includegraphics[width=2\columnwidth]{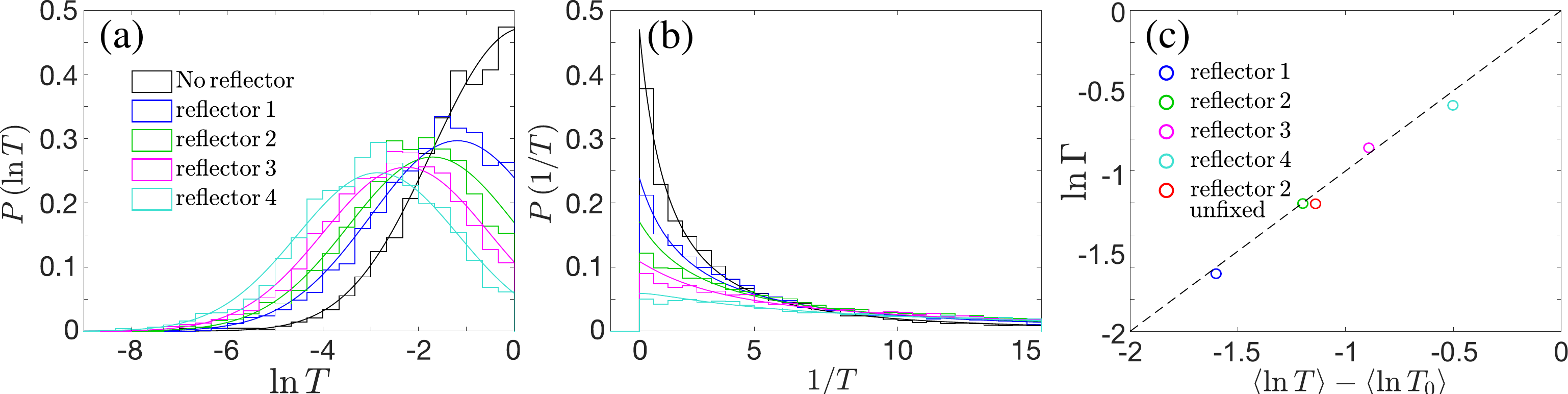}
\caption{(a) Experimental (histograms) and theoretical (continuous curves) of the probability distributions of $\ln T$ with and without a reflector. The scattering strengths of 4 different reflectors are $\ln \Gamma=-0.50,-0.89,-1.20,-1.60$, corresponding to $\Gamma=0.61,0.41,0.30,0.20$. These reflectors are labeled as reflector 1,2,3, and 4, respectively. (b) The corresponding probability distributions of $1/T$. (c) Comparison of the values of the reflector strengths obtained with use of Eq.~(\ref{Gamma}) (y-axis) and with the experimental values $\ln \Gamma=\langle \ln T \rangle - \langle \ln T_0 \rangle$. The red circle corresponds to the result for reflector 2 positioned randomly in different configurations.}
\label{fig_2} 
\end{center}
\end{figure*}

Using the statistical model in the previous section, we average Eq. (\ref{Ttotal}) over the phase $\mu$, to give 
\begin{equation}
\label{invT}
 \left\langle \frac{1}{T} \right\rangle =\frac{1}{\Gamma}\left[ 1 + (2-\Gamma) 
 \langle \lambda \rangle \right]. 
\end{equation}
From the relations $T=1/(1+\lambda)$ and $R+T =1$, we find that $\langle \lambda \rangle=\langle R/T\rangle$. The ensemble average of the ratio of reflection and transmission is given by  $\langle R/T \rangle=(\exp{(2L/\ell)}-1)/2$  \cite{Kumar, Mello1987}, while from Eq.~(\ref{Ttotal}), we obtain:
\begin{equation}
\label{averlnTGamma}
 \langle \ln T \rangle = \ln \Gamma - \frac{L}{\ell},
\end{equation}
where we have used the relation $\langle \ln (1+\lambda) \rangle = L/\ell$ 
\cite{Melnikov,Mello1987}. Using this result, we find    
\begin{equation}
\label{lambdaaver}
 \langle \lambda \rangle =\Gamma^2e^{-\langle \ln T \rangle}-\frac{1}{2}.
\end{equation}
Finally, substituting Eq.~(\ref{lambdaaver}) into Eq.~ (\ref{invT}), 
we obtain the relation between $\Gamma$ and average of functions of the transmission: 
\begin{equation}
\label{Gamma}
 \Gamma = 1-\left[1-\left( 2 \left\langle \frac{1}{T} \right\rangle -1 \right) e^{2 \langle \ln T \rangle}\right]^{1/2} . 
\end{equation}

This result allows us to determine the strength of a reflector inside a disordered medium from transmission measurements.

We measure the transmission spectra from 9.5 to 11 GHz for 50 random  configurations. This gives the ensemble averages required in Eq.~(\ref{Gamma}), as well as the complete  distribution of the transmission. The results of the probabilities distributions
$P \left( \ln T \right)$ and $P \left( 1/T \right)$ are shown in Figs. \ref{fig_2}(a) and \ref{fig_2}(b) for samples with four different embedded reflectors and without a reflector. The histograms and continuous curves correspond to the experimental and theoretical results, respectively . The latter is obtained from Eqs.~(\ref{Ttotal})  and (\ref{poflambda}). For the samples without a reflector, $\langle \ln T_0 \rangle=-1.39$ and the distribution 
of $\ln T$ shown as the continuous black line in Fig.~\ref{fig_2}(a) is similar to a portion of a Gaussian distribution \cite{Muttalib_2003}. As the reflectivity of the fixed reflector increases, the peak in $P \left( \ln T \right)$  moves towards lower values of $\ln T$.

We find the value of  $\Gamma$ by substituting the experimental values of $\langle \ln T \rangle$ and $\langle 1/T \rangle$ in Eq. (\ref{Gamma}). 
Since the logarithmic transmission is an additive quantity, we obtain the experimental values of $\Gamma$ from the difference of the averages of the logarithmic transmission for waveguides with and without the reflector: $\ln \Gamma=\langle \ln T \rangle-\langle \ln T_0 \rangle$. In our experiments, we fix the position of the four different reflectors in the middle of the sample $(x/L=0.5)$ and obtain the following results: $\langle \ln T \rangle-\langle \ln T_0 \rangle=-0.50,-0.89,-1.20,-1.60$, corresponding to samples with $\Gamma=0.61,0.41,0.30,0.20$, for reflectors labeled as 1, 2, 3, and 4, respectively. We note $\Gamma$ can be found in our theoretical model,  Eq.~(\ref{Gamma}), without the knowledge of $\langle \ln T_0 \rangle$, so that measurements of transmission in the samples without a reflector are not needed.
 
The experimental results for the reflector strengths are compare with the theoretical predictions in Fig.~\ref{fig_2}(c), where $\ln \Gamma$, as given by Eq.(\ref{Gamma}),  is plotted vs. the experimental values of $\langle \ln T \rangle-\langle \ln T_0 \rangle$. The results fall close to a straight line with slope 1.  
The red circle in Fig. \ref{fig_2}(c) represents the average over configurations with the reflector 2 positioned randomly in different configurations. The parameter $s$ for the waveguide without a reflector can be obtained from $s=-\langle \ln T \rangle+\ln \Gamma=1.35,1.39,1.42,1.30$ for the ensembles with the four different reflectors. The average magnitude of the difference from  the expected value of $s$ of 1.39 is $3.3\%$.

\begin{figure*}
\begin{center}
\includegraphics[width=2\columnwidth]{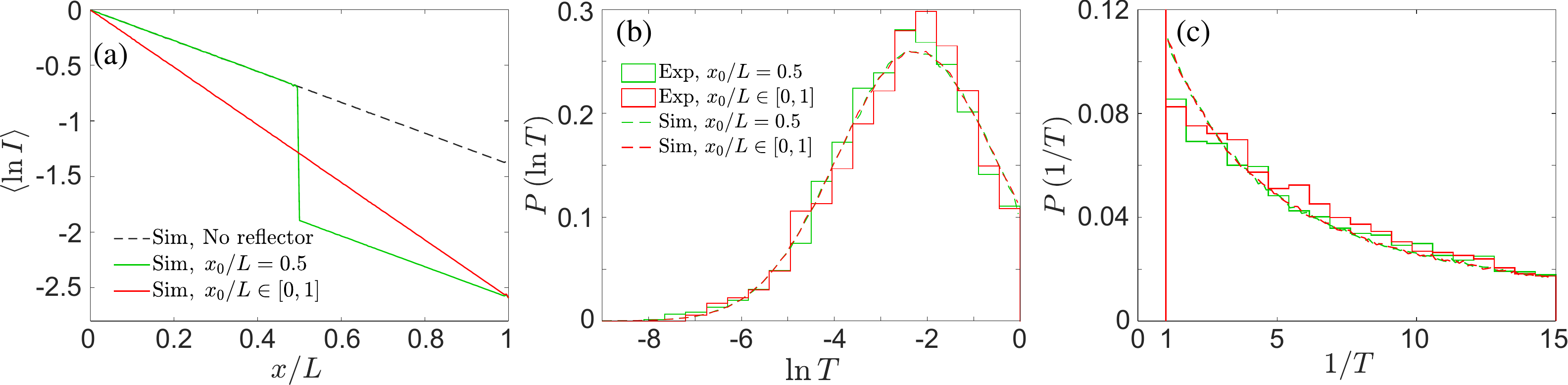}
\caption{(a) Average of the logarithmic intensity $\langle \ln I\left(x\right) \rangle$ from simulations, labeled as Sim, for ensembles without a reflector, with the reflector placed in the middle of the sample, and with the reflector placed in different positions with uniform probability in different configurations. (b) The probability distributions of $\ln T$ for a fixed and a randomly placed reflector. (c) The corresponding probability distributions for $1/T$.}
\label{fig_3} 
\end{center}
\end{figure*}

In accord with the invariance principle, the probability distribution of transmission is independent of the position of the reflector \cite{Huang}. However, the spatial variation of the average logarithmic intensity inside the waveguide $\langle \ln I \left(x\right) \rangle$ naturally changes when the position of the reflector changes with a sudden drop of transmission at the position of the reflector.  This phenomenon is shown in Fig.~\ref{fig_3} using transfer matrix simulations of 1D binary random samples with an embedded reflector of strength $\Gamma=0.41$, as the reflector 2 in the experiments \cite{simulations}. 

A sharp drop of $\langle \ln I \left(x\right) \rangle$ occurs at the position of the reflector, Fig.~\ref{fig_3}(a). But if the average is taken over a random ensemble, in which the position of the reflector equally likely to be in any region of the sample, $\langle \ln I \left(x\right) \rangle$ falls linearly with a slope that would be found for a medium with no reflector but with stronger scattering, such that $s=-\langle \ln T \rangle-\ln \Gamma$. However, the distributions of $\ln T$ and $1/T$ for a fixed reflector and a randomly positioned reflector are the same, as shown in Fig.~\ref{fig_3}(b) and \ref{fig_3}(c), in accord with the invariance principle. In addition, the strengths of reflector 2 evaluated from Eq.~(\ref{Gamma}) for fixed and random positions of the reflector are close, as seen in the closeness of the corresponding green and red circles for reflector 2 in Fig.~\ref{fig_2}(c).

\subsection{Position of the reflector}

\begin{figure}
\begin{center}
\includegraphics[width=1\columnwidth]{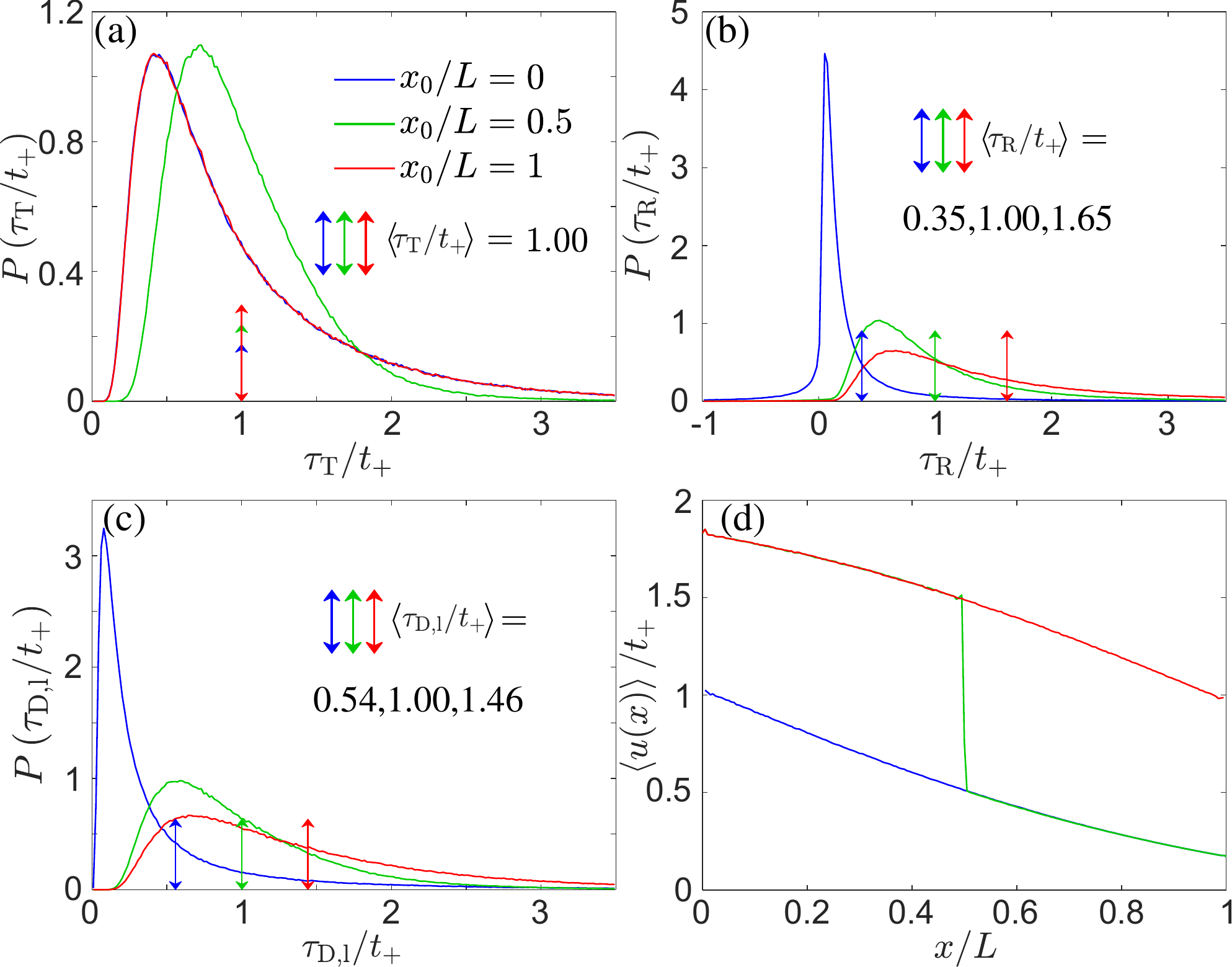}
\caption{Probability distribution of transmission time $\tau_{\rm{T}}$, reflection time $\tau_{\rm{R}}$, dwell time $\tau_{\rm{D,l}}$. The double arrows indicate the values of the ensemble average. (a) Probability distribution of transmission time $\tau_{\rm{T}}$ for 3 different positions of a reflector with $\ln \Gamma=-1.20$. (b) The probability distributions of reflection time, (c) the probability distribution of dwell time and (d) the average of the excited energy $\langle u(x) \rangle$ along the sample for the reflector positions in (a).}
\label{fig_4}
\end{center}
\end{figure}

Though static measurement of transmission cannot yield the position of the reflector, the dwell time of waves incident from one side of the sample, $\tau_{\rm{D,l}}$, allows determining the 
location of the reflector. The dwell time $\tau_{\rm{D,l}}$ corresponds to the energy excited inside the medium by the wave incident from left side of the sample \cite{Arxiv2021,Goto}. This energy varies with the position of the reflector, as can be seen in Fig.~\ref{fig_4}(d). The dwell time is given by
\begin{equation}
\label{tau_D0}
 \tau_{\rm{D,l}}=\int_0^L{u_{\rm{l}}(x)}dx.
\end{equation}
 Here, $u_{\rm{l}}(x)=\frac{1}{2}\epsilon(x)|E_{\rm{l}}(x)|^2$ is the energy density for a wave of unit flux incident from the left.

The dwell time can be also expressed in terms of waves scattered from the sample \cite{Arxiv2021,Goto},  
\begin{equation}
\label{tau_D1}
 \tau_{\rm{D,l}} = T \tau_{\rm{T}} + R \tau_{\rm{R}},
\end{equation}
where  $\tau_{\rm{T}}=d\varphi_{\rm{T}}/
d\omega$, $\tau_{\rm{R}}=d\varphi_{\rm{R}}/d\omega$ and $\varphi_{\rm{T}}$, $\varphi_{\rm{R}}$ are the phases of the transmitted and reflected fields, respectively. The two expressions for $\tau_{\rm{D,l}}$ in Eq.~(\ref{tau_D0}) and Eq.~(\ref{tau_D1}) make it possible to find the position of the reflector from scattered waves.

Measurements of spectra of the phase of the transmitted and reflected fields, were not sufficiently accurate to allow for a determination of the associated times. We therefore carried out numerical simulations of $\tau_{\rm{T}}$, $\tau_{\rm{R}}$ and $\tau_{\rm{D,l}}$  for different positions of reflector 3 with $\ln \Gamma=-1.20$. The results are shown in Fig. ~\ref{fig_4}. The probability distributions of $\tau_{\rm{T}}$ for the reflector at $x_0/L=0$ and 1 are identical since the transmitted field for incident waves from the left and right are identical, as required by reciprocity. This can also be understood by noting that $\tau_{\rm{T}}$ is equal to $\pi\rho$, which is the average of the energy excited by waves incident from both sides of the system  \cite{Arxiv2021}. Because the statistics of total excited energy for waves incident from both sides are the same for reflectors placed symmetrically about the center of the sample, the probability distributions of the transmission time for symmetrically placed reflectors are the same. In addition, the average of the transmission time is independent of the position of the reflector since it is proportional to the average DOS, which is unchanged by the presence of a thin reflector and is equal to the transmission time of wave in the homogeneous medium, $t_{+} =L/v_{\rm{g}}$, where $v_{\rm{g}}$ is the group velocity \cite{Arxiv2021,Razo}.

In contrast to the independence of the average transmission time, the average of the reflection time increases as the reflector moves deeper into the random samples, as would be the case in a sample with uniform dielectric constant.

The reflection time can be very small relative to $t_+$ and in some configurations it may even be negative. This does not correspond to superluminal propagation of information but simply to the reshaping of the scattered pulse \cite{Sebbah}. The dwell time for waves incident from the left, $\tau_{\rm{D,l}}$, is equal to the total excited energy for a wave of unit flux incident from the left, and so is always positive. As the reflector is moved deeper into the sample, the distribution of dwell time for a wave incident from the left broadens and its average value increases. Since $\tau_{\rm{R}}+\tau_{\rm{R}}^{\prime}=\tau_{\rm{D},\rm{l}}+\tau_{\rm{D},\rm{r}}=2\tau_{\rm{T}}$ is valid for each configuration of the disorder, the relation also holds for ensemble average, $\langle \tau_{\rm{R}} \rangle + \langle \tau_{\rm{R}}^{\prime} \rangle = \langle \tau_{\rm{D},\rm{l}} \rangle + \langle \tau_{\rm{D},\rm{r}} \rangle = 2\langle \tau_{\rm{T}} \rangle = 2t_+$. As a result, the sum of the average of reflection time and of the average dwell time for two symmetrical positions about the center is equal to $2t_+$.

More insight into the sensitivity of the dwell time to the location of the reflector can be gained by the impact of the reflector upon the excited energy inside the medium \cite{Huang}, as shown in Fig. \ref{fig_4}(d). The excited energy follows the invariance principle, in which statistical properties of waves are independent of the position of a reflector for any point as long as the reflector is not moved through that point, and the energy density experiences a sharp drop at the position of the reflector. As a result, the total excited energy excited from the left, which is equal to the corresponding dwell time of the wave incident from the left, increases as the reflector moves deeper into the sample.

\begin{figure}
\begin{center}
\includegraphics[width=\columnwidth]{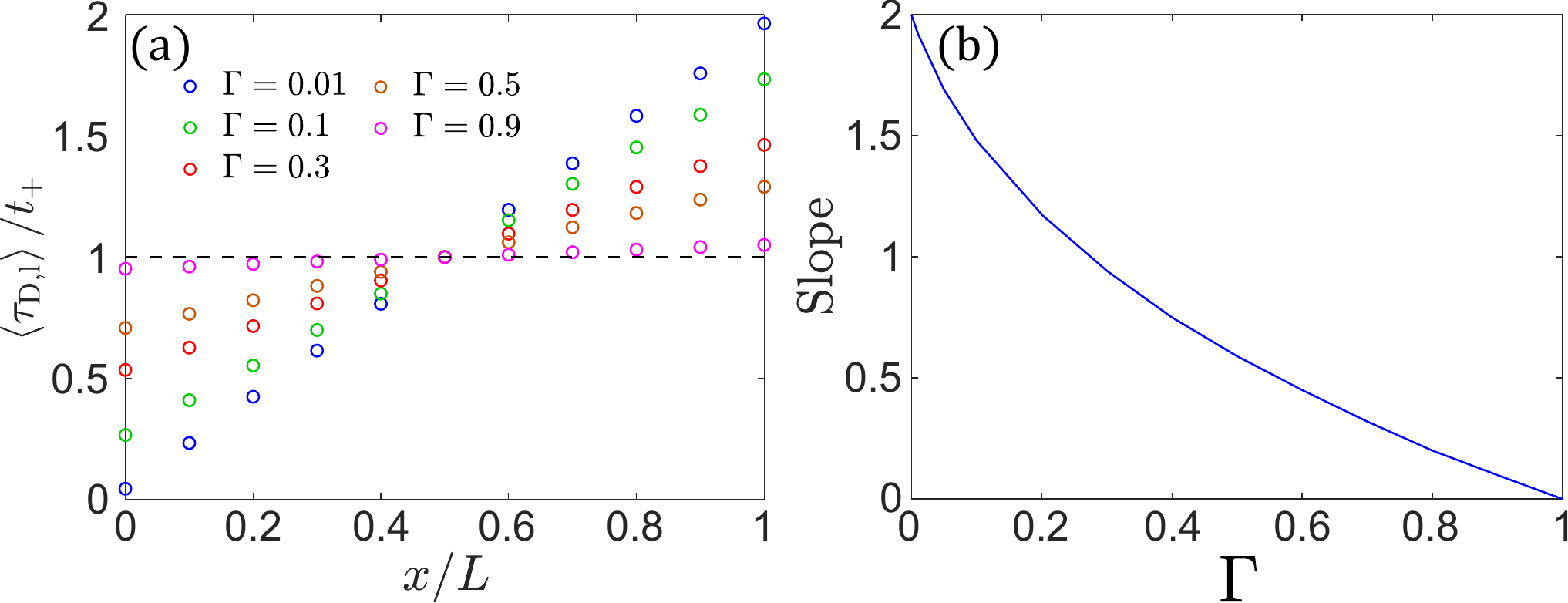}
\caption{(a) Dependence of the average dwell time on the position and strength of the reflector. (b) Slope of the linear relation between average dwell time and the position of the reflector.}
\label{fig_5} 
\end{center}
\end{figure}

We performed numerical simulations of the dependence of the average dwell time upon the scattering strength and the position of a reflector within the medium. The results presented in Fig.~\ref{fig_5} are for a medium with the same value of $s$ as in the experiment, $s=1.39$. The average dwell time $\langle \tau_{\rm{D,l}} \rangle$ increases nearly linearly with the depth of the reflector in the sample. The slope of the variation of dwell time with depth of the reflector decreases as the strength of the reflector decreases, as seen in Fig.~\ref{fig_5}(b). The variation of the dwell time upon reflector position makes it possible to determining the position of a reflector from waves scattered from the sample. 

\section{Conclusions}

The results presented here demonstrate that despite the complexity of wave scattering inside a random medium, it is possible to provide a universal description of wave transport inside a homogeneously disordered 1D medium with an embedded reflector in the sense that statistical results depend only upon the value of $s(=L/\ell)$ of the medium and characteristics of the reflector. In 1D, the reflector is characterized by its reflectivity and position. Furthermore, we show that it is possible to separate the impact of scattering from a fixed reflector and  the homogeneously disordered surrounding medium in scattered waves and to find the mean free path and the transmission coefficient and position of an embedded reflector. The strength of the reflector is given in terms of the averages $\langle 1/T \rangle$ and $\langle \ln T \rangle$, while the position of the reflector is obtained from the average of the dwell time for waves incident from one side of the sample, which can be determined from measurements of the transmitted and reflected field. The solution of this inverse scattering problem for random 1D media and the universality of transport might also be extended to higher dimensions. It may be possible to separately determine the scattering strengths of the bulk medium and the position and strength of a fixed reflector in higher dimensions. 

We have considered a medium without absorption. The presence of uniform absorption can in principle be separated out from other phenomena. The ensemble average of the logarithm of transmission is reduced by absorption and by the presence of a reflector as, $\langle \ln T \rangle=-s-\ln \Gamma-L/\ell_{\rm{a}}$, where $\ell_{\rm{a}}$ is the absorption length, $\ell_{\rm{a}}=v\tau_{\rm{a}}$, $v$ is the wave velocity inside the medium and $\tau_{\rm{a}}$ is the exponential absorption time \cite{Zhang, Arxiv2021}. This suggests that it may be possible to find $\ell_{\rm{a}}$ as well as $\ell$, $\Gamma$, and $x_0$ from radiation scattered from complex systems. 

In this work, the scattering strength of the random sample and reflector were found from the statistics of transmission. In many circumstances, it is not possible to measure transmission. However, the transmission coefficient in lossless systems can be determined from the reflected wave, since $R=1-T$, so that characteristics of the medium and embedded reflected can be found.

\acknowledgements

A. Z. G. acknowledges supported by the National Science Foundation under EAGER Award No. 2022629 and by PSC-CUNY under Award No. 63822-00 51. V. A. G acknowledges support from MCIU (Spain) under the Project number PGC2018-094684-B-C22 and Subprograma Estatal de Movilidad 2013-2016 under the Project number PRX16/00166. He also thanks the Physics Department of Queens College of the City University of New York for their hospitality.

$\dagger$ gopar@unizar.es


\begin{thebibliography}{50}
\bibitem{simulations} We calculate the field inside, transmission and reflection coefficient of a TE polarized monochromatic beam incident on a 1D disordered multilayer sample using transfer matrix formalism.  The sample has a fixed length $L=1$m with the layer thickness  following a uniform distribution in the interval ($0,2d_{\rm{ave}}$), where $d_{\rm{ave}}=0.01$m, and number of layers $N \approx L/d_{\rm{ave}}$. The sample is surrounded by a homogeneous medium with refractive index $n_{\rm{ext}=1}$, and the refractive index of the layers alternates values between the low value $n_1$ and high value $n_2$, with $(n_1+n_2)/2=1$  to minimize the boundary mismatch. The simulation results are averaged over $5\times 10^6$ realizations.
\end{thebibliography}
\end{document}